\documentstyle[twocolumn,prl,aps,epsfig]{revtex}

\begin{document}
\draft

\twocolumn[
\hsize\textwidth\columnwidth\hsize\csname @twocolumnfalse\endcsname

\title{Thermoelectric properties of the degenerate Hubbard model }

\author{V.S. Oudovenko$^{*}$ and G. Kotliar }

\address{ Serin Physics Laboratory, Rutgers University, 136
Frelinghuysen Road, Piscataway, New Jersey 08854, USA}
\date{\today}
\maketitle

\begin{abstract}
We investigate the thermoelectric properties of a system near a
pressure driven Mott-Hubbard transition. The dependence of the
thermopower and the figure of merit on pressure and temperature
within a degenerate Hubbard model for integer filling $n=1$ is
calculated using dynamical mean field theory (DMFT). Quantum Monte
Carlo method is used to solve the impurity model. Obtained
results can qualitatively explain thermoelectric properties of
various strongly correlated materials.
\end{abstract}

]

\narrowtext
The discovery new strongly correlated materials and improvement of
theoretical methods offer new perspectives in the search for systems with
good thermoelectric performance ~\cite{Mahan:1997}. The dimensionless figure
of merit, $ZT={S^{2}\sigma T/\kappa }$, provides us with a quantified
measure of the thermoelectric performance of materials. The higher values of
the thermoelectric figure of merit corresponds to the better thermoelectric
properties. To get maximal thermoelectric response at a fixed temperature, $%
T $, we need to have maximum possible thermopower, $S$, and the electrical
conductivity, $\sigma $, while the thermal conductivity, $\kappa $, should
be the smallest possible. The thermal conductivity has two contributions
electronic, $\kappa _{e}$, and lattice, $\kappa _{L}$, one, $\kappa =\kappa
_{e}+\kappa _{L}$. For some times, the highest value of the figure of merit
was equal to one and only a little improvement in getting materials with
higher figure of merit has been achieved in the last two decades. Since the
electronic structure of strongly correlated electron systems exhibit
properties which have no analogy with those of weakly correlated compounds,
it is important to understand their physical properties and how they impact
the thermoelectric power. Strongly correlated electrons are sensitive to
small changes in their control parameters such as temperature, doping and
pressure and hence require a detailed investigation. There is now a strong
interest in understanding how this changes affect the thermoelectric
properties in order to find more efficient material with higher
thermoelectric response. Even a modest increase in $ZT$ could substantially
impact a number of applications~\cite{Mahan:1997}.

Two important parameters in strongly correlated electron systems
are the carrier concentration and the ratio of the on-site
interaction $U$ to the bandwidth $W$. This ratio can be altered by
applying external pressure or internal pressure by means of
isovalent substitutions.

In this paper we study behaviour of the figure of merit and the
thermopower near a pressure driven Metal-Insulator Transition
(MIT) within framework of the degenerate Hubbard model. \ Our
focus is on the effects of orbital degeneracy. \ There are
several motivations for our study. First, in many strongly
correlated materials the orbitals, which bring a major
contribution into unusual physical properties, are degenerate.
Second, the orbital degeneracy \ allows \ us to study the effects
of particle hole asymmetry \ in the Mott insulating state with
filling $n=1$ in a natural way (one can also introduce particle
hole asymmetry in this case by adding hopping integrals beyond
nearest neighbors). Third, the degeneracy will allow us to use
the integer filling $n=1$ what in the real system corresponds the
situation when the number of the electrons is equal to the number
of sites. We use Coulomb repulsion $U$ and temperature $T$ as the
parameters to vary and take the half bandwidth $D$ to be unity.

While the calculation of the numerical value of the
thermoelectric coefficient requires the detailed modeling of the
band structure and interaction constants of the relevant
compounds, we expect that the {\it qualitative} features related
to the thermoelectricity of a Mott insulator at integer filling
can be captured by the simplest model, the degenerate \ Hubbard
model, and we undertake this study in the paper. \ Indeed the
trends discussed here, are found in ${\rm NiS_{2-x}Se_{x}}$
and ${\rm Ni_{1-x}Co_{x}S_{2}}$ as we discuss toward the end of
this paper. The algorithms necessary for more realistic
calculations of thermoelectricity will be discussed in a future
publication \cite{Palsson:2001}.

To treat the two-band degenerate Hubbard model we use the
Dynamical Mean Field Theory (DMFT) \cite{Review:1996}. Recent
development of the DMFT has given a boost to the study of
strongly correlated systems and their properties. In particular
it has resulted in a detailed understanding of the Mott
transition.

Previous studies of  thermoelectric properties using DMFT were
done within framework of the one-band Hubbard model ~\cite
{Pruschke:1993,Rozenberg:1995,Merino:2000}, strong temperature
and doping dependence of the thermopower was reported. P\'alsson
and Kotliar considered the orbital degeneracy away from half
filling in the limit of infinite interaction 
strength~\cite{Palsson:1998}. A good description of the Seebeck
coefficient dependence on temperature and doping for titanates
was obtained.

In this paper we focus on the integer occupancy case with orbital
degeneracy. We also  keep $U$ finite and study dependence of thermoelectric
properties on the interaction. Changing the ratio of the  interaction
strength one can mimic the effects of pressure.

The $N$-fold degenerate Hubbard Hamiltonian reads:
\begin{eqnarray}
H = && -\sum_{\langle ij\rangle , \sigma } t_{ij}c_{i\sigma}^\dagger
c_{j\sigma} + {\frac{U }{2}} \sum_{i, \sigma \ne \sigma^{\prime}} n_{i
\sigma} n_{i \sigma^{\prime}} -\mu \sum_{i, \sigma } n_{i \sigma},  \label{H}
\end{eqnarray}
where $\langle ij \rangle$ runs over nearest neighbor sites, $\sigma $ is
the spin and orbital index which runs from 1 to $N$. The two-band degenerate
Hubbard model corresponds to $N=4$. The hopping matrix is given by $t_{ij}$,
$U$ is the Coulomb repulsion and $\mu$ is the chemical potential.
We also want to restrict ourselves to the
paramagnetic phase (in both spin and orbit indices) i.e. we are interesting
in metal-insulator transition and its influence on transport properties due
to electronic correlations only. It is worth to mention that orbital
degeneracy suppresses the magnetic correlations enough to make our
assumption easily fulfilled~\cite{Kajuter:1997}. To make computations easier
further we assume $t_{ij}=t$.

In this paper we use the model on a cubic lattice. For calculation of
thermodynamic properties we use the semicircular density of states (DOS) $%
\rho (\epsilon) = (2 / {\pi D}) \sqrt{1 - ( \epsilon /D)^2}$, with the
half-bandwidth $D=W/2=2t$. The semicircular DOS corresponds to an infinite
coordination Bethe lattice which, as it was found earlier\cite{Review:1996},
gives a good description of three dimensional systems.

The next step is to solve the Hubbard model. The standard way in
DMFT to do it is to map the lattice Hubbard model onto the
effective impurity problem, which is a generalized single
impurity Anderson model, where the operators carry an orbital
index. Supplemented with the self-consistency condition~
\cite{Review:1996} solution of the impurity model gives us the
solution of the original Hubbard model. Quantum Monte Carlo
method with extended Hirsch-Fye algorithm
\cite{Hirsh:1986,Takegahara:1992} is used as the impurity solver.
To reduce computational errors (the Trotter breakup) imaginary
time interval is set to $\Delta\tau=1/4$. During the
self-consistency procedure one needs to make direct and inverse
Fourier transformations for the Green's functions (GF). In the
present calculations we use a modified Fourier transformation
(FT) in order to get correct results (for details see Appendix).

The output of the self-consistent procedure described above is
the Green's function on imaginary time axis. To calculate the
thermoelectric properties we need to know behaviour of GF on real
frequency axis. The Maximum Entropy (ME) method is used to make
the analytical continuation of imaginary-time GF to DOS on the
real frequency axis. Knowing imaginary part of GF we reconstruct
frequency dependence of the real part $G(\omega)$ through
Kramers-Kronig relations. Obtained GF is used to calculate the
transport coefficients.

Using Kubo formalism one can express thermoelectric coefficients
in terms of current-current correlation
functions which are reduced within DMFT to averages over the
spectral density function
$\rho(\epsilon,\omega)$~\cite{Palsson:1998,Pruschke:1995}:
\begin{equation}  \label{eq:thp}
S = - {\frac{A_{1} }{e A_{0}}}, \;\;\;\; \sigma = {\frac{e^2 }{T }}A_{0} ,
\;\;\;\; \kappa=(A_2-{\frac{A_{1}^{2}}{A_0}}),
\end{equation}
where the coefficients $A_n$ have the following form,

\begin{eqnarray}
A_{n} &=& {\frac{1 }{V}}\sum_{k,s} \int d \omega \rho_{s}^{2}(k,\omega)({%
\frac{\partial \epsilon_k }{\partial k_x}})^2 (-T{\frac{\partial f(\omega) }{%
\partial \omega}})(\beta\omega)^{n} \nonumber \\
&=&{N_{deg}\pi } \int_{-\infty}^{\infty}d\omega d\varepsilon{\frac{%
\rho^{2}(\varepsilon,\omega) (\omega\beta)^{n}}{4\cosh^{2}({\frac{%
\beta\omega }{2}})}}\Phi(\varepsilon),  \label{eq:An}
\end{eqnarray}
where $T$ is temperature, $f(\omega)$ is the Fermi distribution function, $s$
describes spin and orbital indexes which runs from 1 to $N_{deg}$. The
spectral density, $\rho_{s}(k,\omega)$, and the transport function, $%
\Phi(\varepsilon)$, contain the relevant information about the bare band
structure, $\epsilon_{k}$:

\begin{equation}
\Phi(\varepsilon) = {\frac{1 }{V}}\sum_{k} \left({\frac{\partial \epsilon_{k} }{%
\partial k_x}}\right)^2 \delta(\varepsilon-\epsilon_{k}).  \label{eq:Phi}
\end{equation}
As it is seen from the above formulae (\ref{eq:An}) and (\ref{eq:Phi})
contribution to the thermopower strongly depends on temperature. For low
temperatures only states close to Fermi surface (FS) contribute to
thermoelectric properties while for high temperatures the entire Brillouin
zone is important. Coulomb repulsion, $U$, acts on the transport
coefficients via changes in the spectral density function, $%
\rho(\varepsilon,\omega)$. There is a very simple mnemonic rule
to define the sign of the thermopower. For small temperatures it
depends on DOS slope at the Fermi energy: if the DOS curve rises
when it crosses the Fermi energy then the sign is negative and
vice versa. For very high temperatures it depends on weight
of the Hubbard bands: if weight of the Hubbard band above the
Fermi energy is larger than contribution from the band below the Fermi level
then the thermopower is negative. This rule will help us easily
understand the thermopower behaviour analyzing temperature dependence
of DOS.

In Fig.~\ref{fig:1} we plot dependence of the thermopower on interaction
strength $U$. The thermopower changes sign in a region of Coulomb repulsion $%
2.5<U<3$, the region where the system undergoes metal-insulator
transition at low temperatures. The MIT for high temperatures
transforms into the metal-insulator crossover what is reflected
in the spectral function (DOS) behaviour presented in
Fig.~\ref{fig:2}. It is worth to notice quite strong $S$ dependence
on $U$ in that crossover region. When $U \to 0$
the thermopower will have a finite value depending on temperature and
in opposite limit $U \to \infty$ we expect saturation of
the thermopower dependence which will be also temperature dependent.
The interaction dependence of the thermopower helps us
qualitatively understand the pressure influence onto the system
as well as behaviour of a system (e.g. ${\rm NiS_{2-x}Se_{x}}$)
changing its properties from insulating to metallic ones by
varying the bandwidth. It follows from the fact that the only
important parameter in the system is the the ratio $U/W$ and one
can vary any of two variables to get qualitative description of a
system properties.

Behaviour of the figure of merit on interaction strength
(Fig.~\ref{fig:3}) is possible to understand from its definition.
It has minimum where the thermopower changes sign. After the
sign change it substantially increases and has maximum for higher values of $%
U$. Rather high values of $ZT$ (more than one) can be explained by the
absence of the lattice contribution in the thermal conductivity. If
one takes into account the lattice thermal conductivity the
figure of merit drops to value below one. Usual values of lattice
contribution into the thermal conductivity in transition
metal-oxides lay in region 0.1-10 W/mK. We believe that 1 W/mK is
a reasonable value for pyrites.
We see  that the figure of merit is  maximal for interaction U between
4 and 4.5, but even in this region obtained figure of merit 
could not be competitive with the one for semiconductors
due to the lattice contribution into the thermal conductivity.

Temperature dependence of the thermopower for $U=3$ presented on
Fig.~\ref{fig:4} can be easily explained from analysis of
DOS temperature dependence in Fig.~\ref{fig:5}. For very low
temperatures the thermopower has linear dependence (one can show
it analytically, it is nearly impossible at the present stage to
reach very low temperatures with QMC) and negative sign indicating
presence of quasiparticles in the system what is clearly seen in
Fig. ~\ref {fig:5} ($\beta =16$) where a good quasiparticle peak
at the Fermi energy is observable. With increasing temperature a
smooth crossover from quasiparticles excitations to incoherent
ones reflected in non-monotonic thermopower dependence
which changes sign  from negative to positive one.
In Fig.~\ref{fig:5} for temperatures $\beta =12$ and $\beta =8
$ we see how quasiparticle peak of DOS gets lower and then disappears.
From this observation we can conclude that positive value of $S$
tells us that the band structure consists from two Hubbard
subbands only. Similar behaviour of the thermopower (with the sign change)
occurs in the single band Hubbard model upon
doping~\cite{Pruschke:1995} and in the Hubbard  model with
frustration~\cite{Merino:2000} as well as in the periodic
Anderson model~\cite{Schweitzer:1991}. Further temperature
behaviour of the thermopower is new one.
At temperature $\beta=8$ in Fig.~\ref{fig:4} the thermopower reaches its maximum 
and then starts decreasing (we should notice here that temperature  $\beta=4$ 
is the coherent temperature for the studied system).
For $\beta=4$ the spectral function consists from two bands, where lower
Hubbard band, which is closer to the Fermi energy
has greater  contribution into the thermopower
than the upper one, which is located a bit further from the
chemical potential position. With increasing temperature two
bands became less asymmetric to the Fermi energy and the weight
from the upper Hubbard band ($n=1$) is larger than the
contribution from the lower band. It means that the thermopower
should became negative as it seen in Fig.~\ref {fig:4} (the last
point). And finally, for very high temperature the two
bands collapse into the one.
But the sign of the thermopower remains unchanged by the
same reason as we discussed above. The thermopower sign change 
in the  high temperature limit due to Mott-Hubbard bands
collapse was reported in ~\cite{Yao:1996}. 

To gain further understanding of the temperature and interaction
dependence of the thermopower we study its high-frequency
behaviour in the high-temperature limit. To do so we generalize the
thermoelectric response to finite frequencies
(ac-thermopower)~\cite{Mahan}:

\begin{equation}  \label{Sw}
S(\omega ) = - \frac{1}{{eT}}\frac{{L^{12} (\omega )}}{{L^{11} (\omega )}},
\end{equation}
where coefficients $L^{11}$ and $L^{12}$ are defined as
\[
L_{ji }^{1n} (\omega ) = \frac{{ie^{\beta \Omega } }}{{\omega \beta }}
\sum\limits_{\mu \nu } {} \frac{{\left\langle \mu \right|q_i \left| \nu
\right\rangle \left\langle \nu \right|J_j ^n \left| \mu \right\rangle (e^{ -
\beta \varepsilon _\mu } - e^{ - \beta \varepsilon _\nu } )}} {{\omega +
\varepsilon _\mu - \varepsilon _\nu + i\delta }}.
\]
Here $e^{ - \beta \Omega } = Tr\left( {e^{ - \beta (H - \mu N)} } \right)$, $%
i,j = x$, $J_x^1 = j_x$, $J_x^2 = j_x^Q $, $j_x$ and $j_{x}^{Q}$ are
electrical and heat currents in $x$ direction correspondingly, $q_x$ is the
polarization operator satisfying $j_x(t)=\partial q_x (t)/\partial t$.

Expanding numerator and denominator of the thermopower ($L^{11}(\omega)$ and
$L^{12}(\omega)$) in frequency and dividing one onto another we obtain the
following expansion of the thermopower in the high-frequency limit:

\begin{equation}  \label{Swcomm}
S(\omega ) = - \frac{\beta }{e}\left( {\frac{{\left\langle {\left[ {q_x
,j_x^Q } \right]} \right\rangle }}{{\left\langle {\left[ {q_x ,j_x } \right]}
\right\rangle }} + {\cal O(\omega )}} \right) .
\end{equation}

The relevant commutators are given by:

\begin{eqnarray}
& & - \frac{1}{it }[q_x ,j^{Q}_{x} ] = \\
& &\sum\limits_{js } \{ t[c_{j + 2x}^\dag c_{js} + c_{j - 2x, s}^\dag
c_{j,s} ] + \mu [c_{j + x,s }^\dag c_{j,s} + c_{j - x,s}^\dag c_{j,s}]
\nonumber \\
& & - U[c_{j + x,s}^\dag c_{j,s} \, \sum\limits_{s^{\prime}\ne s} {n_{j +
x\,s^{\prime}} } + c_{j - x,s}^\dag c_{j,s} \,\sum\limits_{s^{\prime}\ne s}
n_{j - x\,\,s^{\prime}} ] \} .  \nonumber
\end{eqnarray}

This expression consists of three  terms. The first one, proportional to $t$%
, doesn't contribute to the high-temperature expansion if one takes into
account only nearest neighbour hoppings. The second term (proportional to
chemical potential) similar to what we have in the denominator and the last
one proportional to $U$ contribute to high-temperature expansion of the
thermopower. The final result  we end up with is presented below
\begin{equation}
S(\omega )\stackrel{T\rightarrow \infty }{\longrightarrow }-\frac{\beta }{e}{%
\left( {-\mu +U\frac{\bar{n}}{1-n}}\right) },  \label{mainresult}
\end{equation}
where $\bar{n}$ corresponds to contribution to filling $n$ coming from $%
N_{deg}-1$ degrees of freedom (spin and orbital) and is equal to $%
\sum_{\sigma ,\sigma ^{\prime }\neq \sigma }\langle (1-n_{\sigma
})n_{\sigma ^{\prime }}\rangle $. If one assumes high-temperature
behaviour of the chemical potential in the form $\mu /T=\alpha $,
where $\alpha $ should be negative for the filling
$n<{\frac{1}{2}}N_{deg}$, one obtains that the sign of the
thermopower should be negative in the high-temperature limit.
Hence  ac- and dc- thermopowers have the same sign in the
high-temperature limit.

It is quite difficult to make a direct comparison of a model
calculation and situation in real materials due to complexity and
wide variety of real structures. To make a comparison with our model
calculations we need a material with degenerate $e_g$ band hosting
one electron (hole). As we mentioned in the introduction a good candidates for the
comparison are pyrites compounds with doubly degenerate
$3d-e_{g}$ Ni band. An attempt to understand high-temperature
dependence of the thermopower in this materials on base of
one-band electron correlation theory with a good fit of
experimental data using six parameters model was given some years
ago in work of Kwizera et al.~\cite{Kwizera:1980}. To explain low
temperature dependence of the thermopower it was suggested that
both holes and electrons participate in charge transport and a
two-band model was proposed as an appropriate model for data
interpretation~\cite{Yao:1996}. Our calculations show that the
two-band degenerate Hubbard model treated within DMFT, is
sufficient model for at least qualitative description of pyrites
in the whole temperature range~\cite
{Yao:1996,Kwizera:1980,Mabatah:1977}. In the case of half-filled
Ni $3d-e_{g}$ band in ${\rm NiS_{2-x}Se_{x}}$ theoretical model under consideration
 due to the symmetry in the system should give zero thermopower for all temperatures,
while experimentally it is zero only for temperature $T<100$~K.
For higher temperatures the thermopower becomes positive. It is clear
that one needs to take into account  effects of other bands
(filled $t_{2g}$ band in the first turn). The thermopower
behaviour in ${\rm Ni_{1-x}Co_{x}S_{2}}$ with temperature looks
rather similar as in  ${\rm NiS_{2-x}Se_{x}}$  but with Co
substitution of Ni atoms occupancy of $3d-e_g$ band changes from
two ($x=0$) to one ($x=1$). One can track the
experimental situation in ${\rm NiS_{2-x}Se_{x}}$ analyzing model
calculations (for temperatures low enough to neglect contributions coming
from other bands) especially in the case ${\rm CoS_2}$ which
corresponds to one-quarter filled $3d-e_g$ band.

In conclusion, we calculated dependencies of the thermopower and
the figure of merit on interaction strength and temperature of
the thermopower in the two-band degenerate Hubbard model for
integer filling $n=1$. The strong dependence on studied parameters
was obtained. Analytical estimations of the high-frequency limit
of ac-thermopower is provided. In the high-temperature limit ac-
and dc- thermopowers have the same sign (negative for filling
$n<0.5$). We expect that the thermopower behaviour in the
three-band degenerate Hubbard model would behave in a similar
way. It means that we can qualitatively understand behaviour of
the thermoelectric properties in a wide range of strongly
correlated materials where a major role played by $d$- and $f$-shell
electrons.

Systems near the temperature driven Mott transition exhibit very
rich thermoelectric behaviour. As we cross the
localization-delocalization threshold the thermopower increases
substantially. There are the high-temperature precursors of the
first order Mott transition that takes place at lower
temperature. As the temperature is lowered, and the critical Mott
endpoint is approached, the specific heat diverges and the entropy
jumps. The figure of merit is low in this crossover regime
because of the great deal of cancellation between the
quasiparticle contributions which are in our model electron like,
and the Hubbard band contributions which are hole like. A
detailed phase diagram temperature versus filling for different
values of $U$ using the current approach will be studied
elsewhere.

The authors are indebted to G. P\'{a}lsson, A.I. Lichtenstein for many
helpful and stimulating discussions. We also acknowledge usage of Rutgers
Computational Grid (RCG) PC cluster as well as NERSC Cray T3E supercomputer
which made our computations possible. The research was supported by American
chemical society Petroleum Research Fund grant ACS-PRF \# 33495-ACS.

\appendix

\section{Fourier transformation in QMC calculations}

In the self-consistent procedure to solve the impurity problem
using QMC method~\cite{Review:1996} we need to do two direct
Fourier transformations for GFs $G(\tau )$ and $G_{0}(\tau )$
(the Weiss function) and one inverse Fourier transformation (iFT)
for $G_{0}(\tau )$ only. Function $G_{0}(\tau )$ is an input for
QMC simulations producing $G(\tau )$. Two direct Fourier
transformations are necessary to impose the self-consistency
condition which is usually written in frequency space. Both, GF
and the Weiss function contain the discontinuities at $\tau =0$
and $\tau =\beta $. The discontinuities in these functions and
their  derivatives determine the high-frequency behaviour
of their frequency dependent Fourier transformations $G(\omega )$ and $%
G_{0}(\omega )$. As it is well known that information about high-frequency
behaviour of GFs is absent in QMC simulations themselves (maximum frequency
available is the Nyquist frequency $\omega _{max}=1/2\Delta $, where is $%
\Delta $ is imaginary time interval). \ \ The high frequency (small
imaginary time) information is not contained in the QMC
itself but has to be incorporated into the splining procedure using
additional information.

This information is available from calculation of corresponding moments of \
the GFs. \ To make \ this connection clear, we make consecutive integration
by parts of the Fourier integral:
\begin{eqnarray}
G(i\omega_n)&=&\sum_{k=0}^{N} \frac{(-1)^{k+1}(G^{(k)}(0)+G^{(k)}(\beta))}{%
(i\omega_n)^{k+1}} \\
&+&\frac{(-1)^{N+1}}{(i\omega_n)^{N+1}} \int\limits_{0}^{\beta} {\rm {e}}%
^{i\omega_{n}\tau} \frac{\partial ^{N+1} G(\tau)}{\partial \tau^{N+1} }
d\tau .  \nonumber  \label{eq:GFF}
\end{eqnarray}
Values of GF sum and it's derivatives can be expressed via corresponding
momenta. These momenta can be calculated similar to the one-band Hubbard
model ~\cite{Nolting:1997}. A momentum of $k$ degree is defined as follows:
\begin{equation}
M^{(k)} = \int\limits_{-\infty}^{+\infty} d\omega \omega^{k} {\rho}%
_{}(\omega) .  \label{eq:Mn}
\end{equation}

We can bind Eqs. (A1) and (\ref{eq:Mn}) writing the following expression for
sum of GFs and it's derivatives in imaginary-time space:
\begin{equation}
G^{(k)}(0)+G^{(k)}(\beta )=M^{(k)},  \label{eq:GM}
\end{equation}
where $k=0,\ldots N$. 
To make direct Fourier transformation, first, we interpolate $G(\tau )$
defined in $L$ points ($L$-number of time slices) and then take analytical
Fourier transformation of resulting function. The procedure of interpolation
plays a  key role. We used the cubic spline interpolation for $G(\tau )$
where condition of continuous second derivative $(G^{(2)})$ is imposed.
Using this condition we write a system of linear equations to find
interpolation coefficients. To close the set of linear equations we need to
put correct boundary conditions. The standard way to do it is to set the
second derivatives at the end points of the time interval $[0,\beta ]$ to
zero (so called, natural spline). In our approach we use analytical
information about the momenta (sum of the second and first derivatives at
the end points, $M^{(1)}$ and $M^{(2)}$).

To obtain this information we need to know expansion of the self-energy in $%
1/\omega $ series or in another words we need to know first two momenta of
the self-energy.
\begin{eqnarray}
M_{\Sigma }^{(0)} &=&(2N-1)Un,  \label{sigma:moments} \\
M_{\Sigma }^{(1)} &=&{(2N-1)U^{2}n(1-n)+2U^{2}C_{2}^{2N-1}\left\langle {nn}%
\right\rangle },  \nonumber \label{A4}
\end{eqnarray}
where $N$ is the number of bands and $C_{n}^{k}=\frac{(k)!}{n!(k-n)!}$ is a
combinatorial factor which arrives due to the \ spin and orbital degeneracy.

The self-energy expansion contains density-density correlations
functions for different spins and orbitals. One way to obtain
them is to use an approximate scheme which is accurate at high
frequencies such as the coherent potential approximation (CPA)
~\cite{Velicky:1968}, \cite {Kajuter:1996}. Another\ possibility
is to evaluate the correlation functions which enter in
Eq.~(A4) using  the  QMC procedure \ which allows us to
compute arbitrary local correlation functions such as the
density-density correlators. The self-consistency procedure
ensures us that they are correct correlation functions
when the self-consistency is
reached. Having correct momenta we solve the set of linear
equations to obtain the cubic spline function which is Fourier
transformed analytically.

The power of modern computers allows us to use in imaginary time ($\tau$)
space up to $L=256$ time slices only, while in frequency ($\omega$) space we
do not have such limitations and can use as much frequency points as we want
but only frequencies less than the Nyquist frequency have physical meaning
(usually it is less than $2^{8}$). As we know GF has $1/\omega$ asymptotic
behaviour and dealing with the finite number of frequency points we
introduce the finite energy cut-off. Cutting off the tail of GF, we remove
the discontinuity from the Fourier transformed function in $\tau$ space. To
correct the situation we subtract the high-frequency tail from GF and make
FT of the obtained function numerically and make Fourier transform of the
tail analytically. Finally, in $\tau$ space we sum up the obtained functions
and as the result we have the correct inverse Fourier transformation, we
call it the ``new" one.

To demonstrate the difference between two ``old" and ``new" Fourier
transformations we calculate the self-energy in the considered model for $%
U=4 $ and $\beta=8$. Results are plotted in Fig.~\ref{fig:6} where
the ``new" solution is drawn by solid line and the ``old" one is
plotted by dashed line. We see that the ``old" one has the region
where the self-energy changes sign (``overshoots"). It
corresponds to unphysical contribution to self-energy which
should always keep the same sign for positive or negative
frequencies.

Finally, we stress that difference between two Fourier transformations
becomes substantial especially in critical regions of parameters: low
doping, high values of $U$ and low temperatures $T$.

\newpage

\begin{figure}
\caption{ Dependence of the thermopower, $S$ ( in units of $k_B/e =86
\mu V/K$ ), on Coulomb interaction, $U$ (in $D=1$~eV) for
$\beta=8$.} \label{fig:1}
\end{figure}

\begin{figure}
\caption{Spectral functions (DOS) for $U=2.5$ and 3 for
$\beta=8$.} \label{fig:2}
\end{figure}

\begin{figure}
\caption{ The figure of merit, $ZT$, vs.   interaction strength,
$U$, for temperature $\beta=8$ and different values of the
lattice conductivity $\kappa_L$~[W/mK].} \label{fig:3}
\end{figure}

\begin{figure}
\mbox{}\\[-1.5cm]
\caption{Temperature dependence of ther\-mo\-po\-wer, $S$ for
$U=3$. The solid line is guide for the eye.} \label{fig:4}
\end{figure}

\begin{figure}
\caption{ Temperature dependence of density of states for
$U=3$.} \label{fig:5}

\end{figure}

\begin{figure}
\caption{Dependence of imaginary part of the self-energy  on
Matsubara  frequency axis $\omega$ ($\beta=8$) for two different
Fourier transformations `old' (dashed line) and `new' (solid
line) one.} \label{fig:6}
\end{figure}

\newpage

\begin{figure}
\centering
\includegraphics[height=5cm,angle=0]{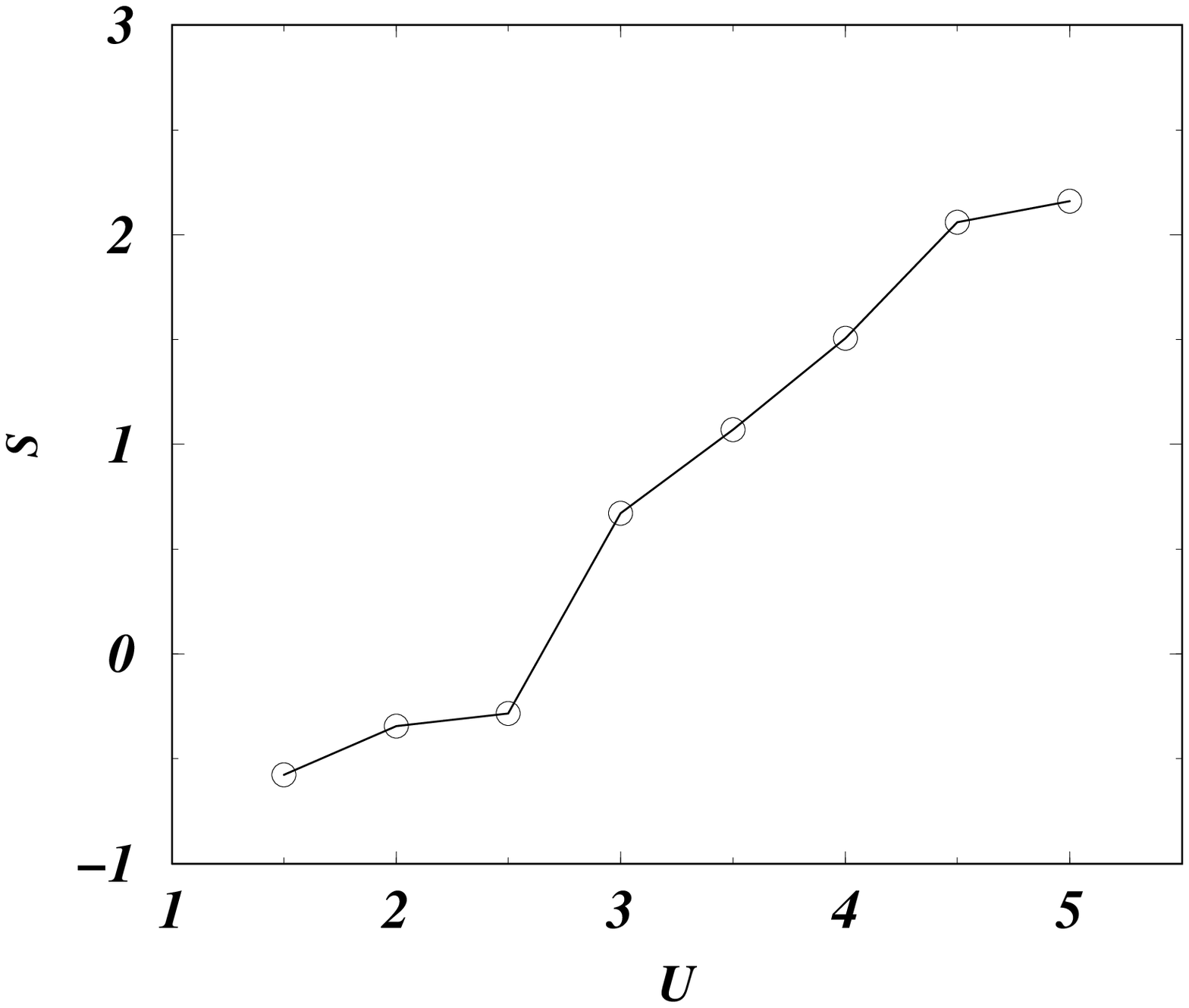}\\
FIG. 1
\end{figure}

\begin{figure}
\centering
\includegraphics[height=5.5cm,angle=0]{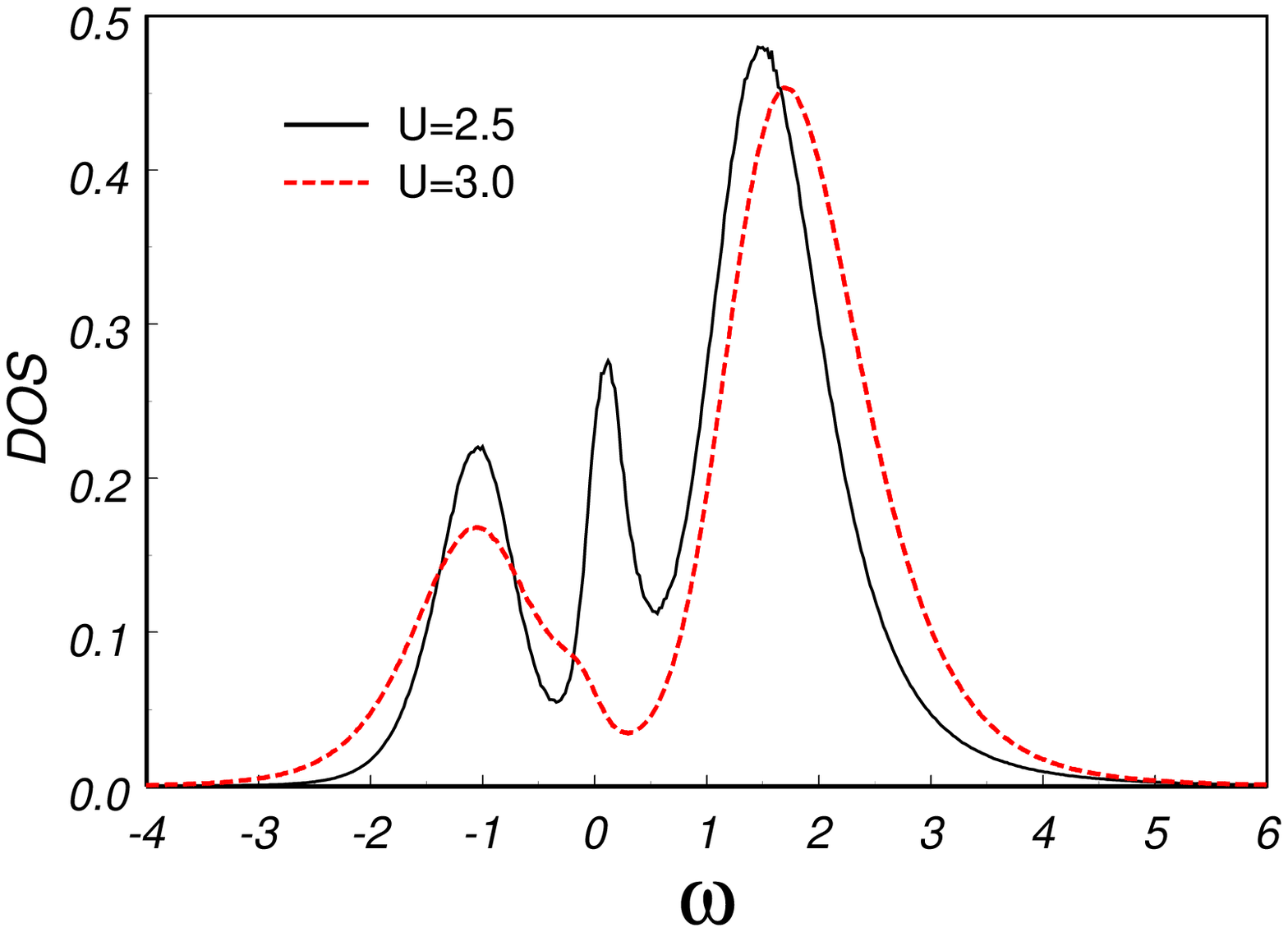}\\
FIG. 2
\end{figure}

\begin{figure}
\centering
\includegraphics[height=5.5cm,angle=0]{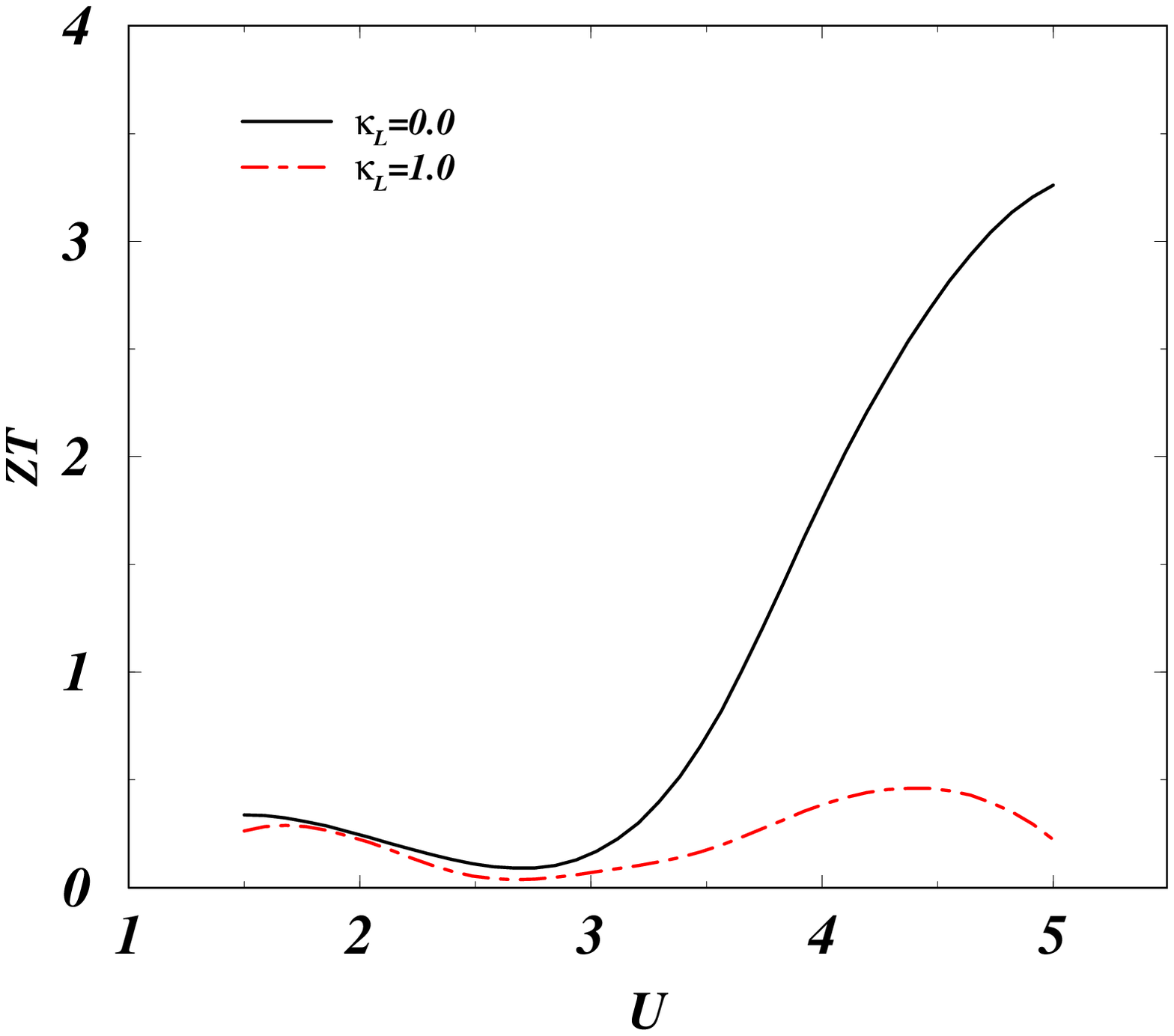}\\
FIG. 3
\end{figure}

\begin{figure}
\mbox{}\\[-1.5cm]
\centering
\includegraphics[height=6cm,,angle=0]{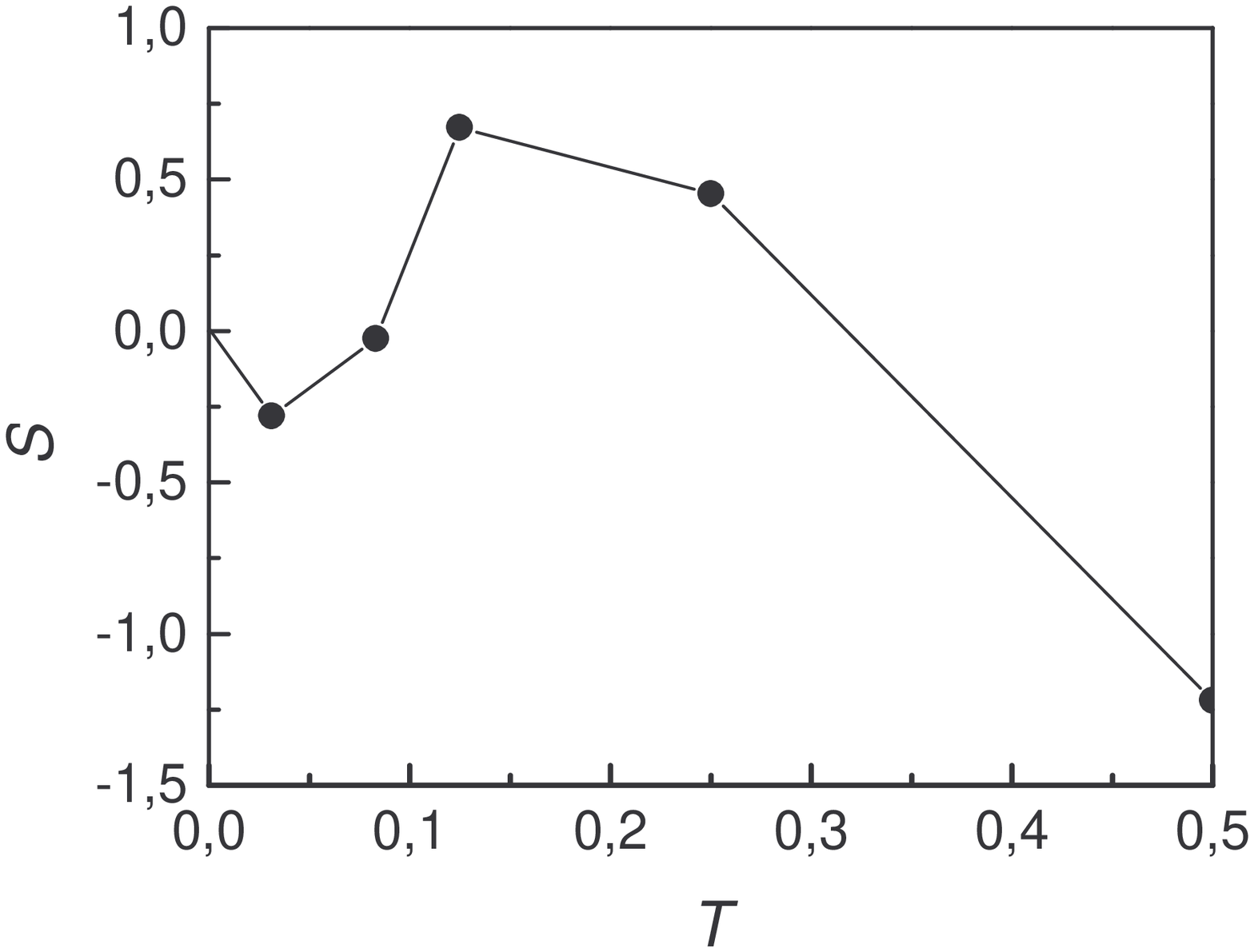}\\
FIG. 4
\end{figure}

\begin{figure}
\centering
\includegraphics[height=4.5cm,angle=0]{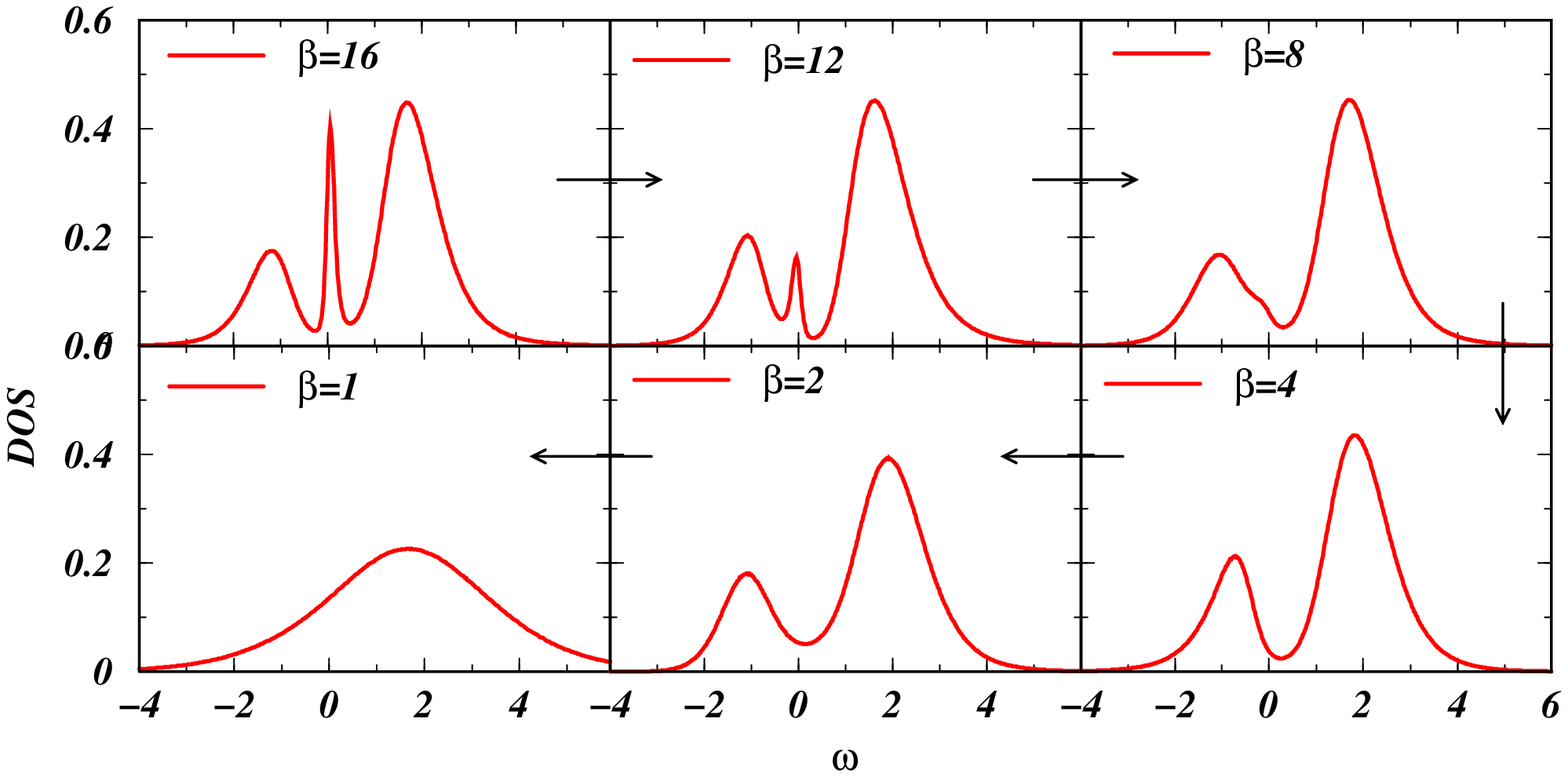}\\
FIG. 5
\end{figure}

\begin{figure}
\centering
\includegraphics[height=5.5cm,angle=0]{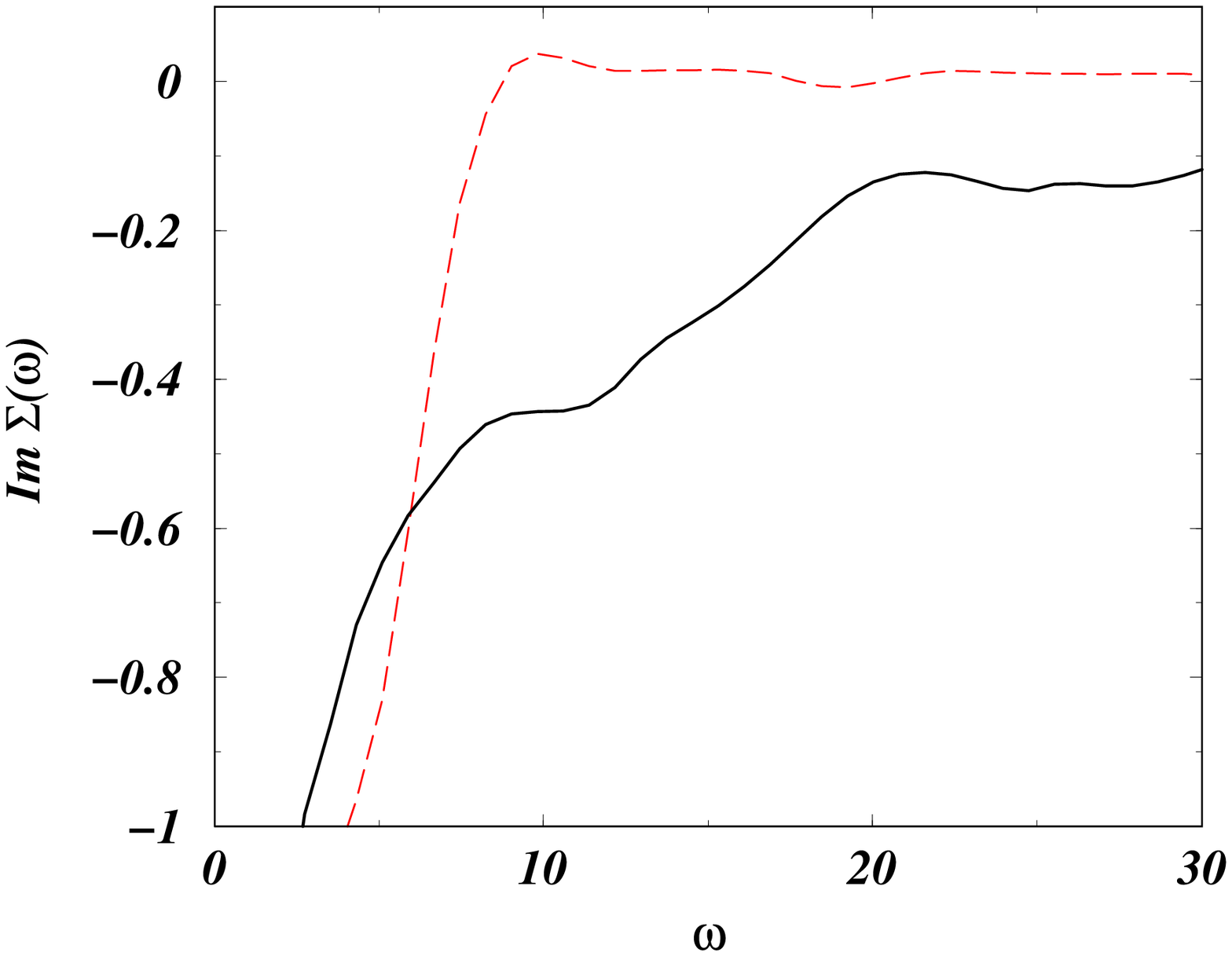}\\
FIG. 6
\end{figure}

\end{document}